*This is the accepted version of the Ethics of Open Data chapter (4) of the Open Data Section. NO FURTHER CHANGES WITHOUT EDITOR APPROVAL!*

# Ethics of Open Data

Brandon Locke & Nic Weber

# Introduction

This chapter addresses emergent ethical issues in producing, using, curating, and providing services for open data. Our goal is to provide an introduction to how ethical topics in open data manifest in practical dilemmas for scholarly communications and some approaches to understanding and working through them. We begin with a brief overview of what can be thought of as three basic theories of ethics that intersect with dilemmas in openness, accountability, transparency, and fairness in data: Virtue, Consequential, and Non-consequential ethics. We then map these *kinds* of ethics to the practical questions that arise in provisioning infrastructures, providing services, and supporting sustainable research in science and scholarship that depends upon open access to data. Throughout, we attempt to offer concrete examples of potential ethical dilemmas facing scholarly communication with respect to open data, and try to make clear what kinds of ethical positions are helpful to practitioners. In doing so, we hope to both clarify the ethical questions facing librarians doing practical work to support open data access, as well as situate current debates in the field with respect to these three kinds of ethics.

# Open Data and Ethics: A Brief Overview

Ethics can be practically framed as "the study of the general nature of morals and of the specific moral judgments or choices to be made by a person."[1] This definition situates ethics as a matter of individual choice, but of course the choices we make as individuals have broad impacts on the communities we are part of, serve, and wish to see flourish. That is, ethics is often practically framed as the result of individual choices and actions, but ethics also encompasses the implicit and explicit values of an institution, community of practice, or even group of researchers. The relationship between individual choice and collective action is particularly relevant for scholarly communications where curators, librarians, and policy makers collaboratively work to provide regular and unfettered access to resources needed to conduct research, develop guidelines or regulations that govern ethical behavior, and practically implement standards that encode or formalize these rules. It is important to acknowledge at the beginning of this chapter that morals and judgements, whether individual or collective, don't arise from the ether - they are grounded in beliefs about what is right, just, or serves the greater good given an alternative set of choices. The ethical dilemmas faced by a community of practitioners are often about deciding what is right, how justice is enacted and preserved, or what choices we make to produce the greatest good for the communities on whose behalf we work.

Broadly speaking, ethics are often broken down into an answer to three questions about **how** to make the "right" or correct choice:

---

[1] Stuart A. Burns, "Evolutionary pragmatism, a discourse on a modern philosophy for the 21st century." *The purpose of ethics* (2012), http://www3.sympatico.ca/saburns/index.htm.

1. Virtue ethics: *What action best exemplifies a high moral character?*[2]
2. Consequential (or Teleological) ethics: *What action best achieves desired consequences?*[3]
3. Non-Consequential (or Deontological) ethics: *What action best adheres to established rules*?[4]

In the following subsections we explain how these questions can be reframed and made specific to the context of scholarly communication and open data.

## Virtue Ethics

Virtue ethics focus on the moral character of the action or policy. An approach to open data through a virtue ethics standpoint may say, "Broader access to more information is inherently good for the scholarly community. How can we provide access to our research materials to the most people possible?" While this perspective remains amenable to personal interpretations and understandings, there are a few different aspects one may pursue.

For example, consider data access and reuse justice: Do some instances of data sharing privilege particular people or groups, or does anyone who is interested have free and equal access?

When data is published freely online, anyone with sufficient technology and internet access is free to view and evaluate the data that was produced in the course of research or other data collection processes. When data is not published openly online, some may gain access to data through paywalls, or professional networks, but many others would not. Gatekeeping in this manner, even when unintentional, means that the data would be used only by people with personal connections, or people who have the time and knowledge to make a request. Open licensing, as one alternative, could allow individuals interested in the data to not have to seek out explicit permission to use or modify the data, which likewise removes a barrier to usage that is often more difficult for people without a connection to the creator or an impressive institutional affiliation. Further embrace of open research methods, such as non-proprietary formats, clear documentation, open source software, and open code minimizes barriers to data access, and adheres to principles of virtue ethics that seek fair and just access and use to all potential users.

Some questions faced by those working under a virtue ethics umbrella in scholarly communication include:

---

[2] Rosalind Hursthouse and Glen Pettigrove, "Virtue Ethics", *The Stanford Encyclopedia of Philosophy* (Winter 2018 Edition), Edward N. Zalta (ed.). https://plato.stanford.edu/archives/win2018/entries/ethics-virtue/.

[3] Walter Sinnott-Armstrong, "Consequentialism", The Stanford Encyclopedia of Philosophy (Summer 2019 Edition), Edward N. Zalta (ed.). https://plato.stanford.edu/archives/sum2019/entries/consequentialism.

[4] Larry Alexander and Michael Moore, "Deontological Ethics", The Stanford Encyclopedia of Philosophy (Winter 2020 Edition), Edward N. Zalta (ed.). https://plato.stanford.edu/archives/win2020/entries/ethics-deontological.

- What is the just and fair way to provide access to data?
- What are the rights of human participants, and what are their reasonable expectations for being notified about the use of their personally identifiable information?
- What are the moral obligations of researchers in sharing data as a common pool of resource that can be expediently reused?

# Consequential (Teleological) Ethics

Consequential ethics is often framed around utilitarian notions of how the greatest net benefit can be achieved through individual and collective action. A consequential approach to open data would weigh the potential benefits of making data available against the potential harms of doing so. There are many arguments for greater openness providing the greatest good from the creation and curation of data, but data openness may also bring potential harm.

## Scholarly Research Data

In academia, many argue that open data derives benefits for the creator of the data, the health of a scholarly community, and for the efficient and fair reuse of data.[5] Creators of open data see greater scholarship impact metrics, such as increased citations and downloads on research articles and associated datasets, on open access materials than on paywalled or privately held research materials (e.g. Piwowar and Vison, 2013, Christensen et al., 2019, Colavizza et al., 2020). The scholarly community benefits by encouraging reproducibility, and allowing other scholars to critique or verify results of empirical studies. Other researchers benefit from open data by having a greater number of resources to build upon in effectively designing and conducting research.

## Government, Nonprofit, and Corporate Data

There are also many benefits to data openness outside the academy. Government agencies (local, state, and federal level), NGOs, and corporations often make some amount of data open and available to the public to encourage transparency and economic, social, and political innovations. The citizens, patrons, and partners of these organizations also value open data because it can be used (at least in theory) to hold organizations accountable for the work they're doing, and develop trust with stakeholders.[6] Another benefit of these types of open data is that the data collected can be used by community members, including businesses and nonprofits, enabling more organizations and individuals to glean insights into the area without the expense, time, and (in some cases), intrusiveness of collecting the data on their own.[7] In this context, open data can benefit the community or communities that are being represented, the organizations that are making the data available, and the public at large.

---

[5] Laure Perrier, Erik Blondal and Heather MacDonald, "The views, perspectives, and experiences of academic researchers with data sharing and reuse: A meta-synthesis," *PLOS ONE* 15(6): e0229182 (2020); https://doi.org/10.1371/journal.pone.0229182.

[6] Marijn Janssen, Yannis Charalabidis and Anneke Zuiderwijk, "Benefits, adoption barriers and myths of open data and open government," *Information systems management* 29:4 (2012): https://doi.org/10.1080/10580530.2012.716740.

### Increasing Benefits and Minimizing Harms

From one perspective, the more open and accessible the data is, the more potential benefit it could bring. In other words, data licensed under CC-BY or similar schemes would enable the largest potential reuse possibilities, without deterring uses that may be commercial or mandating that any users use the same open license. Putting data into the public domain, or using a CC0 license could garner even greater impact among user communities, though it may ultimately be detrimental to the dataset's creator by potentially reducing citations.

However, any harm caused by the data's release would negate the benefits derived from making data openly accessible and available. This harm may be enacted through a number of different ways—personally identifying information may be improperly disclosed, or data that highlights a particular marginalized community could allow people to further marginalize that community. Individuals may also feel negatively about the process if they are not properly informed of the data collection, or if they see the data reused in ways that they did not expect or approve of. These harms are not evenly distributed across society, and largely exacerbate existing inequality and injustice.

Thus, a consequentialist's approach to open data would weigh likely and potential benefits and harms, trying to determine which modes of access create the greatest overall good given the consequences that can be predicted.

Some questions posed by a consequentialist view of ethics are:
- How do we create the greatest utility from data? How can research investments be maximized in terms of credit and benefit that accrue to those that are the subjects of data, and those that are the stakeholders of the results (e.g. the general public)?
- What steps (e.g: de-identification and anonymization, licensing, metadata and documentation) can be taken to minimize harm while still maximizing benefits?
- Which standards enable the efficient and effective reuse of data? How should curators or scholarly communication librarians advocate for and promote the use of open formats, encodings, and even analytic software to promote efficient reuse?

## Non-consequential (Deontological) Ethics

Non-consequential, or deontological ethics, usually frame morality around adherence to norms, laws, or other authority structures (e.g. compliance with rules established by scholarly societies, codes of ethics). There are many established layers of laws, rules, and best practices concerning the sharing of research data. These stem from laws protecting privacy and intellectual property, funder mandates, institutional rules, and best practices dictated by the scholarly community. Some of these may limit the information you can collect and share, while others may require that you share things openly (though exceptions are usually built into public access requirements).

---

[7] Tim Davies and Zainab Bawa, "The Promises and Perils of Open Government Data (OGD)," *The Journal of Community Informatics* 8 (2), (2012); http://ci-journal.net/index.php/ciej/article/view/929.

## Privacy Laws

When it comes to human subjects, the ways data are collected and shared may be limited by broad federal laws in the locations where subjects live. One of the most well known among these is the US [Health Insurance Portability and Accountability Act (HIPAA)](#), which has a Privacy Rule that outlines individuals' rights to control the collection and sharing of their health information. Another impactful law is the European Union's [General Data Protection Regulation (GDPR)](#), which requires anyone who handles human subject data to design systems with privacy in mind, and requires informed consent and privacy rights to individuals.[8] At the current moment, several other countries and a few US states are considering legislation modeled after GDPR, potentially greatly broadening overarching human subject protections.

## Open Access Mandates

When research is funded by a governmental or private entity, funding bodies may have rules mandating that research outputs, including research data, be made publicly available. In 2013, the White House Office of Science and Technology Policy mandated that all federal granting agencies develop a plan to make research publicly available within 12 months of publication.[9] As a result, federal agencies include requirements for data sharing in their terms and conditions documentation. Likewise, many private funding agencies may also require that research materials are published openly, and will include details in grant documentation.

## Institutional Review Boards

If work is being conducted at or in collaboration with a college or university and constitutes human subjects research, research plans (including what data are and are not made publicly available) are subject to the institution's Internal Review Board (IRB), which can approve, require modifications, or disapprove research based on the potential for harms.

---

[8] For more on the impact of GDPR on research data, see Phillips, M., & Knoppers, B. M. (2019). Whose Commons? Data Protection as a Legal Limit of Open Science. *The Journal of Law, Medicine & Ethics*, *47*(1), 106–111. https://doi.org/10.1177/1073110519840489.

[9] White House Office of Science and Technology Policy, I*ncreasing Access to the Results of Federally Funded Research*, Memorandum for the Heads of Executive Departments and Agencies, Washington, DC. 2013, https://obamawhitehouse.archives.gov/sites/default/files/microsites/ostp/ostp_public_access_memo_2013.pdf.

Although IRB approval is sometimes misunderstood as the be-all and end-all of research ethics, that is simply not true. While they are an important part of an ethics review, reviews may differ greatly on different campuses, and many historical cases of research malfeasance have been approved by IRB. In addition, IRBs have a fairly narrow scope, and do not generally cover data collection that doesn't include interactions with human subjects, such as public records, remote sensing, or information accessible on personal blogs or social media. Research that uses these types of sources, particularly blogs and social media, can clearly have a negative impact on the safety, well-being, and privacy of people included (see Case 2). Many professional organizations also have best practices and guidelines for data sharing and privacy protections within the relevant research domain, which may offer guidelines that are better suited for the research than IRB, or may cover ethics in areas broader than the IRB scope.

### Limitations of Existing Laws and Norms

It is important to note that there are many other potential harms that fall beyond these different sets of established rules. No set of laws, rules, and norms could cover all potential harms, and researchers and data curators should participate in further processes regarding the potential benefits and harms an open dataset could cause. These processes should best be created in conversation with the community or communities that are most represented or impacted by the dataset, as well as with other scholars or ethicists.

Some questions asked from a non-consequentialist perspective of open data ethics include:
- How do data management practices conform with requirements from funding agencies?
- What laws and regulations (local or national) govern the sharing of data? What role should curators play in enforcing and upholding these rules?

## Synthesis of Different Ethical Perspectives on Open Data

What makes ethics interesting and difficult in the context of open data, and scholarly communications more generally, is that the choices between one ethical perspective and another are often about differences in *degree* rather than *kind*. In other words, ethical choices related to open data are about balancing virtue, utility, and rule conformance. As practitioners, we rarely have the liberty to make a wholesale choice between, say, utility and virtue. Instead, we are forced to weigh trade-offs between, for example, how much utility can be achieved in providing access to data without compromising things like the privacy rights and expectations of research participants. An example of how these three ethical schools intersect is made clear in questions such as:

- What licenses or clauses in data regulations [non-consequential] can guarantee data are maximally beneficial for reuse [consequential], and how should they be enacted fairly [virtue]?
- Who deserves access, why, and under what circumstances? This includes both the *duties* of those collecting, curating, or managing access to data as well as the *rights* of

> individuals that have collected data, and critically those that are the subjects of data collection and the norms and *expectations* of public scholarship, civic governance, etc.

These choices are context-dependent and largely a matter of the institutional setting that practitioners find themselves.

# Case Studies

In the remainder of this chapter we explore specific cases or topics in open data, and attempt to unpack the normative assumptions behind actors that may be involved. For each case reviewed, we first describe the topic, the stakeholders, and the type of data that are involved. We then attempt to explain ethical trade-offs made, and how these impact access, privacy, intellectual property, or even data quality.

## Case 1: Alice Goffman's 'On the Run' Ethnography

Some of the most sensitive data generated in the course of empirical research comes from the close study with and of people in their built environments. Observational work that takes seriously the notion of ethnomethodology is, for example, often rooted in practices of reflexivity and participant engagement that places the researcher as a steward of private (personal) information. Increasingly, calls for transparency across the spectrum of qualitative scholarship (e.g. Elman and Kapiszewski, 2014) encourage the open sharing of instruments and data used to generate novel findings when research is conducted with and through participant observation.

In this setting, Alice Goffman's book 'On the Run: Fugitive Life in an American City' (2015) became a crucible for numerous ethical controversies in contemporary qualitative research, ethnography, and sociology. The data collected for Goffman's book came from over 6 years of intensive field work in which Goffman not only observed, but lived in a neighborhood of Philadelphia where she befriended a group of black men eking out a living amongst unjust systems of oppression and lack of social or economic mobility. The book is, regardless of the controversy surrounding its production, an incredible commentary on the vicious cycle that perpetually ensnares young black men in the USA's criminal justice system. As part of her fieldwork, Goffman attended bail hearings, court dates, and observed the harassment of her subjects by police officers as they went about navigating life in a Philadelphia neighborhood.

Where Goffman came under ethical fire in academia was for her positionality in relation to crimes that she observed and arguably participated in. 'On the Run' details a number of episodes in which a crime may have been committed, and many argued that Goffman crossed a line from simply 'participant observation' with her subjects and became implicit in actions that were criminal.[10] The documentary evidence of these controversial events were recorded in

---

[10] Steven Lubet, *Interrogating ethnography: Why evidence matters*. (New York: Oxford University Press, 2018).

Goffman's ethnographic fieldnotes - which after publication - became highly sought after by scholars (as well as law enforcement) looking to substantiate claims that seemed, on their face, difficult to square with official records of crimes, hospitalizations, and reported locations of her research participants. In both anthropology and sociology ethnographic fieldnotes are rarely shared between researchers or made publicly available for secondary analysis.[11] Controversially, Goffman destroyed all of her fieldnotes upon completing the project. The destruction of this data was justified as a way to protect participants, and shut a door to future criminal investigations.

## Ethics at Stake

Part of what has made the controversy of Goffman's work so intense and salient for scholarly communications comes from the notion that a researcher places themselves as the final arbiter in determining what should be done with information collected during the course of a research project. While institutional review boards (IRBs) are often helpful at initially determining risk and credentialing researchers for the ethical management of data, IRBs are not, solely, responsible for the continued management and safeguard of sensitive information when a research project ends. In fact, some IRBs encourage researchers to destroy data to minimize institutional risk upon the end of a sanctioned data collection and analysis period.[12]

Critics of Goffman's destruction of her empirical data do so through a consequentialist framework - arguing that, instead of adhering to principles of transparency and research accountability, she violated her ethical duty to provide evidence that could support claims made in her book. In a normative ethical framework there are many additional questions about the alternative decisions that were available to Goffman. For example, working with data curators she might have, instead of destroying these fieldnotes, created a dark archive of her research field notes which would embargo access for an extended period of time (e.g. 25 years and beyond). Doing so would allow scholars in the future to access her work after legal statute of limitations expired. In taking this route Goffman would have been making a tradeoff that had consequential ethics (achieving the best outcome for ensuring the validity of her findings) and virtue ethics (which would have privileged the protection of her research participants). But, doing so requires trust that such an archive could readily be protected from law enforcement's inquiry, and that there were sufficient long-term preservation policies in place to ensure that a substantial amount of effort in curating this data could be sustained over time. There are, in fact, few examples of how this 'dark archive' proposal may have been practically achieved. In the USA (where Goffman is located) there is only one data repository solely dedicated to the curation of qualitative research data (the Qualitative Data Repository) which has only been in operation since 2014. One can hardly fault Goffman for not being up to speed on alternative choices available, and in some sense adhering to the best practices recommended by an IRB.

---

[11] Victoria Reyes, Victoria. "Three Models of Transparency in Ethnographic Research: Naming Places, Naming People, and Sharing Data." *Ethnography* 19, no. 2 (June 2018): https://doi.org/10.1177/1466138117733754.

[12] Jessica Mozersky, Heidi Walsh, Meredith Parsons, Tristan McIntosh, Kari Baldwin and James M. DuBois, "Are we ready to share qualitative research data? Knowledge and preparedness among qualitative researchers, IRB Members, and data repository curators," *IASSIST Quarterly* 43(4):952 (2020), https://doi.org/10.29173/iq952.

For information professionals, the ethical choices available for managing and preserving data may seem clear and straightforward - but practically speaking the ability to identify and pick out institutional mechanisms that can preserve one's data remains a challenge. In Goffman's case there was a disciplinary legacy, IRB policies, and research positionality that impacted ethical decisions around data access. Each of these factors influenced a decision to destroy data that could and might have been otherwise preserved and made accessible into the future.

## Case 2: Facebook Emotional Contagion Study

The second case study of open data ethics that we explore in this chapter is related to the collection of human subject data that is purported to be open and free for research use. In 2014 researchers at Facebook conducted a study in which they manipulated the news feeds of 700,000 users of the online social networking platform.[13] The goal of this work was to understand the impact of post contents on the users reported 'well being' (in social psychology the concept of well-being relates to their 'moment to moment emotional experiences' and overall life-satisfaction). The authors of this study were, in some sense, trying to refute early studies of social comparison - which holds that the emotion of news posts correlate inversely with an end-user's feelings of well being. That is, if a news post is positive then a user will report negative well-being (e.g. the feeling of being "alone together"), and negative posts invoke feelings of compassion and solidarity (e.g. misery loves company). The study demonstrates that even with 'positive' news the end-user can experience negative and positive emotions - that the relationship between post sentiment and end-user well being is not as directly (negatively or positively) correlated as has been previously reported. This study is unique, however, in that the authors are not depending upon self-reports, but instead mining social media logs of users to infer their sentiment and well-being.[14] The controversial nature of this mining is well reported, but what sparked outrage was the finding that by invoking negative feelings (that is, the researchers injected negative emotional content into a user feed) a Facebook user can experience acute levels of low well-being. This means that, in a sense, the Facebook researchers were invoking negative well-being in users for the sake of conducting their research. Research interventions that cause harm to participants are often very closely regulated by an IRB, and strict controls are supposed to be in place for such procedural research. However, it was unclear whether or not these researchers had taken such steps. In post-publication reporting, there were numerous conflicting reports about whether or not IRB clearance had been obtained, how the interventions were controlled, and what if any outreach to harmed participants occurred.[15]

---

[13] Adam D. Kramer, Jamie E. Guillory and Jeffrey T. Hancock, "Experimental evidence of massive-scale emotional contagion through social networks," *Proceedings of the National Academy of Sciences* 111 (2014): https://doi.org/10.1073/pnas.1320040111.

[14] Timothy Recuber, "From Obedience to Contagion: Discourses of Power in Milgram, Zimbardo, and the Facebook Experiment," *Research Ethics* 12, no. 1 (January 2016): https://doi.org/10.1177/1747016115579533.

In an interview after publication the editor of the article - when asked about its contribution to 'well-being' studies - stated that, "I think it's an open ethical question. It's ethically okay from the regulations perspective, but ethics are kind of social decisions."[16] Indeed, the Common Rule - which governs medical and behavioral research - states that not only should research participants be informed of the studies they are participating in (and facebook users were not informed of their participation) but that researchers should take care not to actively knowingly cause harm if it does not outweigh the significance of the findings of the research (broadly construed).

Ethically, what is at stake in the post-hoc analysis of the Facebook emotional contagion study is a confluence of all three ethical positions. From a Facebook user's perspective - the purposeful manipulation of how posts are promoted and displayed to their social network violates assumed virtue of the platform; users have a common and justified assumption that their personal posts are treated fairly by the service provider - when that assumption is manipulated for research purposes it would seem to violate both an implicit norm (that they are treated equally on the platform) and explicit norm (a 'common rule' that governs ethical behavioral research). When both of these norms are violated then ethically the data shouldn't be considered purely open, but that user data was co-opted and manipulated to meet a narrow research purpose (as argued by Selinger & Hartzog, 2016). From a consequentialist perspective, this study could be ethical if the collected data - whether publicly available or not - resulted in findings that were highly valuable to a research phenomenon. But, many have argued that the co-opting and manipulation of users taints the research findings - and that interpretations that might be gleaned are therefore of little value when relationships between service providers, users, and researchers are contested.[17] Finally, from a deontological perspective of adhering to rules - it remains, as we stated earlier, unclear from multiple ethical viewpoints whether or not this research conformed with ethical obligations of informed consent. Surely, users who agree to terms-of-service agreements that are lengthy and do not clearly explain the reuse of their personal data are unaware that they may be, unwittingly, enrolled in social experiments. These ethical issues are complex in the context of social media platforms, user expectations, and research data collection. Where norms are assumed, and contexts of good will are violated - its hard to argue that data, even when accessible, is reasonably "open" for researchers to ethically use. That the experimental conditions require the manipulation of users on the Facebook platform further throws into doubt whether or not the findings can and should be trusted.

---

[15] Evan Selinger and Woodrow Hartzog, "Facebook's Emotional Contagion Study and the Ethical Problem of Co-Opted Identity in Mediated Environments Where Users Lack Control," *Research Ethics* 12, no. 1 (January 2016): https://doi.org/10.1177/1747016115579531.

[16] Adrienne LaFrance, "Even the Editor of Facebook's Mood Study Thought It Was Creepy," *The Atlantic*, 2014.

[17] Blake Hallinan, Jed R Brubaker, and Casey Fiesler, "Unexpected Expectations: Public Reaction to the Facebook Emotional Contagion Study," *New Media & Society* 22, no. 6 (June 2020): https://doi.org/10.1177/1461444819876944.

# Case 3: Future of Privacy Forum's City of Seattle Open Data Risk Assessment

## Context

While this chapter is primarily focused on scholarly research in the university, it can also be helpful to examine other open data contexts. The City of Seattle in Washington launched its Open Data Program in 2010 and stepped up their intentional publishing of open data in 2016 with an [Executive Order](#) making all City data "open by preference." This Executive Order set the expectation that all city departments will make their data accessible to the public after taking steps to protect the privacy of those represented in the data.

From fall 2016 through summer 2017, The Future of Privacy Forum (FPF), a data privacy advocacy group, conducted an in-depth privacy and risk assessment in collaboration with the city, and published an [extensive report](#) in January 2018. The report uses a blend of virtue, non-consequential, and some consequential ethics, along with practical technologies and processes, to offer rich opportunities to understand the kinds of risks associated with open data, demonstrate risk mitigation techniques and methods to balance both access and privacy, and model the type of ongoing assessment of risks that are necessary for researchers and data curators.

## Data Harms

The report found that there are three main harms that may arise from the sharing of open data: re-identification, data quality and equity, and loss of public trust.

**Re-identification** is when data scrubbed of identifying traits such as names or addresses is still able to be used to determine the identity of a person. This can occur through the use of other additional datasets, or through a combination of data points that are unique to a specific individual. Preventing re-identification can be difficult, particularly as data brokers provide more data and stronger re-identification tools are developed. Data providers should regularly conduct audits and tests to see if re-identification is possible when appropriate.

**Data quality and equity** are flaws in the data collection and/or design that can lead to inaccurate, unintended, or unjust outcomes due to the use of open data by governments and other organizations. The report notes that organizations increasingly rely on public data, and that data is often consumed and repurposed, meaning that errors and biases may have both immediate and long-term effects. Data quality and equity harm can only be reduced through regular, in-depth assessments of published data.

**Public trust** can be damaged when individuals feel their privacy has been violated, or that community expectations are disregarded or otherwise unmet. This can lead to lower participation or false data out of fear of harm caused by a lack of privacy or other protections. Public trust can be protected through clear communication to individuals affected about how data will be collected and shared, and what steps are taken to protect privacy.

## Practices and Process

The Future of Privacy Forum report investigated and scored a wide range of factors in the way the City of Seattle handled their open data pipeline across six domains: privacy leadership and program management, benefit-risk assessment, de-identification tools and strategies, data quality, equity and fairness, and transparency and public engagement. In doing so, they raised a number of details worth highlighting:

### Post-Publication Review

The report gave the City credit for having a fully documented and implemented pre-publication benefit-risk assessment at the time of the review, but found that these assessments weren't regularly reviewed after publication, and the City didn't have any implemented procedures that would trigger re-assessment.

### Training

People within different departments of the city have some privacy governance responsibilities, in addition to Open Data and Privacy programs in the IT department. There was also broad training, even for non-technical employees, to help them identify and understand privacy policies and potential failures. The report noted that this makes data more likely to be protected throughout the full lifecycle of collection, use, release, and disposal.

#### Social Equity

The City has an established [Race and Social Justice Initiative](Race and Social Justice Initiative) program dedicated to eliminating racial disparities in Seattle, and the report noted that the Open Data Program has committed to releasing open datasets that help with promoting positive RSJI outcomes. While the report does not detail how these programs may work together, it is essential to keep in mind that marginalized communities face greater risks from open data privacy breaches and algorithmic decision-making[18]. The inclusion of programs like this, or other experts or specialists can greatly mitigate data harms.

### Outreach and Feedback

The report notes that there is a significant amount of community outreach, including social media, public hackathons, presentations to community groups, and community design workshops, many of which were hosted by local community groups, businesses, and academic institutions. In addition to fostering awareness and use of the available data, this also helps to identify errors or gaps in the dataset. The report did also note that the City has not expended much effort to provide notification to the community about data collection or the possibility of the data being released openly, outside of the Privacy Statement.

### Benefit-Risk Analysis

The city has used a benefit-risk analysis to determine if and how datasets should be shared. This assessment isn't available in the report, but FPF recommended they conduct the assessment yearly for each dataset, and provided an example for future use. Using 5 point scales of "Very Low Likelihood" to "Very High Likelihood" of occurrence and "Very Low Impact" to "Very High Impact" of foreseeable risks, the model benefit-risk analysis provided by the FPF asks that the city:
- Evaluate the information the dataset contains
    - Evaluate any data that may be identifiable, including limiting indirect identifiers, sensitive attributes, and data that is difficult to de-identify
    - Consider how linkable the information in the dataset is to others
    - Consider how the data was obtained to determine accuracy and if publishing would violate any trust
- Evaluate the benefits associated with releasing the dataset
    - Determining which communities will likely use the data and the likelihood and impact of foreseeable benefits
- Evaluate the risks associated with releasing the dataset
    - Consider the foreseeable privacy risks of the dataset, including the impact of re-identification and false re-identification on individuals and the City
    - Consider the foreseeable risks regarding the data quality and equity, including how it may reinforce biases, adverse or discriminatory impacts, and unequal distribution of harm
    - Determine if public trust would be harmed due to public backlash, individuals' or communities' shock or surprise about being included, chilling effect on business or communities, or reveal nonpublic information about an agency's operations

---

[18] There is a rapidly increasing body of work in this area, most notably, Noble, Safiya Umoja (2018), *Algorithms of Oppression*, NYU Press, New York, NY, Hoffmann, Anna Lauren (2019) "Where fairness fails: data, algorithms, and the limits of antidiscrimination discourse", Information, Communication & Society, 22:7, 900-915, DOI:10.1080/1369118X.2019.1573912, and Eubanks, Virginia (2017) *Automating Inequality: How High-Tech Tools Profile, Police and Punish the Poor*, St Martin's Press, New York, NY.

The scores are then tabulated for each dataset to determine if it should be open, have limited access, if the city should conduct additional screening, or if it should not be published. Finally, countervailing factors that may justify releasing a dataset openly regardless of privacy risk are evaluated and taken into consideration.

The Future of Privacy Forum report and risk assessment methodology provide an excellent model for the types of evaluations and assessments open data providers must routinely perform in order to protect the privacy of human subjects and reduce the potential for inequitable outcomes from their work. While there are certainly differences in terms of accountability, legal requirements, and data production and publication between civic data and scholarly research data, this process serves as a model for anyone participating in the collection and curation of open data.

# Emerging Issues in Open Data Ethics

In the concluding section of this chapter we point to future directions and emerging issues related to the curation, use, and library services around open data. We mention briefly two areas where increased access to data poses challenges to ethical data use: privacy and data responsibility.

## Privacy

One of the inherent tensions in data access is protecting privacy and providing unfettered opportunities for meaningful reuse. A number of privacy preserving techniques for releasing data are beginning to see real world implementations. For example, census data collected by the USA provides some of the most valuable demographic open data used by researchers. The 2020 Census will be accessible to researchers through the use of a privacy preservation technique called 'differential privacy' - which allows for computation against data without actually obtaining or storing data on local machines.[19] This is the first and by far the broadest implementation of differential privacy for open data, and time will tell if this is an effective method for protecting sensitive data from harmful reuse. Data trusts are also emerging as a sociotechnical intervention for effectively sharing public and private sector data for research purposes.[20] Data trusts rely on third party intermediaries storing and governing access to data based on researcher credentials. The data trust model is yet to be implemented at scale, but promising experiments such as the University of Washington's [Transportation Data Collaborative](#)—hosting rideshare, bikeshare, and transportation data from municipal bus service providers—are at early stages of development. Additionally many data repositories are offering increased service to help researchers adhere to data sharing requirements and simultaneously protect human participants. ICPSR for example offers a "researcher passport" that credentials users through a screening process that allows for graduated access to restricted data.[21] Similarly the Qualitative Data Repository offers services and protocols for researchers managing privacy sensitive information about human subjects.[22]

---

[19] Alexandra Wood, Micah Altman, Aaron Bembenek, Mark Bun, Marco Gaboardi, James Honaker, Kobbi Nissim, David R. O'Brien, Thomas Steinke, and Salil Vadhan. "Differential privacy: A primer for a non-technical audience." *Vand. J. Ent. & Tech. L.* 21 (2018): 209.

[20] Kieron O'Hara, "Data trusts: Ethics, architecture and governance for trustworthy data stewardship," 2019, https://eprints.soton.ac.uk/428276/1/WSI_White_Paper_1.pdf.

While many challenges remain, each of these advances in technology and the curation of open data attempt to provide a way for access to be enhanced while minimizing risk to the rights and expectations of research subjects.

## Data Responsibility

Increasingly data about vulnerable populations are used to advance policy agendas and advocacy positions. These data are some of the most fraught and complicated resources that require strict safeguards and community protocols for responsible use. The [US Indigenous Data Sovereignty Network](), for example, is an organization dedicated to advancing the governance of indigenous data. US IDSN prompted indigenous data sovereignty, which is defined as "the right of a nation to govern the collection, ownership, and application of its own data." Technological advances for respecting and promoting data sovereignty are emergent, but not without precedent. The Exchange for Local Observations and Knowledge of the Arctic (ELOKA) project is one of the earliest and most advanced examples of community-informed data collection policies and local data storage which respects indigenous communities contributions to and ownership of data.[23] More conceptually, in 2020 D'Ignazio and Klein published [Data Feminism](), a work extolling the virtues of data protection through the lens of care-work.[24] This book provides a logic for operating from and with primary concern for data subjects through the rejection of traditional collection paradigms, as well as analysis techniques which often gloss over the positionality of research subjects.

# Implications for Information Professionals

Information professionals should keep data ethics at the core of their research data management services. There is never a bad time to remind community members to be thinking about the ethics of their research! Ethical data collection, storage, and access should be integrated into scholarly communication and data curation instruction. Developing strong working relationships and open lines of communication with subject librarians (particularly in areas with a lot of sensitive or personally identifiable information like health and social science) can also help to integrate ethical research and data practices throughout the institution. Data management consultations are another good time to discuss ethics with researchers—while discussing data management plans, information professionals can help to create ethical data practices at the start of a research project. Of course, not all researchers prioritize ethics or consult early on data management, so information professionals who deposit or ingest research materials into an institutional repository can also serve as the final line of protection before sensitive material gets published.

---

[21] Margaret C. Levenstein, Allison R. B., Tyler and Johanna Bleckman, "The Researcher Passport: Improving Data Access and Confidentiality Protection," 2018. http://hdl.handle.net/2027.42/143808.
[22] Sebastian Karcher, Dessislava Kirilova, and Nicholas Weber, "Beyond the Matrix: Repository Services for Qualitative Data," *IFLA Journal* 42, no. 4 (December 2016): https://doi.org/10.1177/0340035216672870. ; Karcher, S., & Weber, N. (2019). Annotation for transparent inquiry: Transparent data and analysis for qualitative research. *IASSIST Quarterly*, *43*(2). https://doi.org/10.29173/iq959
[23] Peter Pulsifer, Shari Gearheard, Henry P. Huntington, Mark A. Parsons, Christopher McNeave and Heidi S. McCann, "The role of data management in engaging communities in Arctic research: overview of the Exchange for Local Observations and Knowledge of the Arctic (ELOKA)," *Polar Geography* 35 (2012): https://doi.org/10.1080/1088937X.2012.708364.
[24] Catherine D'Ignazio and Lauren F. Klein, *Data Feminism*. (Boston:MIT Press, 2020).

One other space where data ethics comes into play for information professionals is in crafting data collection policies for their own institutions. Libraries are increasingly collecting and using data, or are providing data to vendors through licensing and access agreements. While this is rarely *open* data, many large library vendors aggregate data from many different places to sell to third parties like marketing firms or law enforcement.[25] The Digital Library Federation's [Privacy and Ethics in Technology Working Group](#) is one notable group working to better understand how patron data is being used, and to better equip librarians to protect patron privacy. The group has produced a number of valuable resources, including "[A Practical Guide to Performing a Library User Data Risk Assessment in Library-Built Systems](#)" and "[Ethics in Research Use of Library Patron Data: Glossary and Explainer](#)."[26]

# Conclusion

In this chapter we have reviewed three ethical positions from which data curators, scholarly communications librarians, and researchers often operate: virtue, consequential, and non-consequential (or deontological). In practice, these ethical decision-making frameworks often intersect—requiring careful attention to not just the moral imperative of protecting harm but also attempting to build technologies and services which can create a greater research good. Through three case studies we have attempted to make clear how ethical decisions complicate the way data is made accessible and are used for research purposes. By describing what is at stake when ethical decisions are faced by stewards of data we believe there is an opportunity to continue improving not only the openness of data, but the moral grounds upon which such decisions are made by individuals and institutions engaged in promoting open research practices.

---

[25] Sarah Lamdan," Librarianship at the Crossroads of ICE Surveillance," *In the Library with the Lead Pipe*, 2019: http://www.inthelibrarywiththeleadpipe.org/2019/ice-surveillance/ & SPARC 2021 https://sparcopen.org/news/2021/addressing-the-alarming-systems-of-surveillance-built-by-library-vendors/

[26] Kristin Briney, Becky Yoose, John Mark Ockerbloom, and Shea Swauger, "A Practical Guide to Performing a Library User Data Risk Assessment in Library-Built Systems," Digital Library Federation, May 2020, http://doi.org/10.17605/OSF.IO/V2C3M; Andrew D. Asher, Kristin A Briney, Gabriel J Gardner, Lisa J Hinchliffe, Bethany Nowviskie, Dorothea Salo, and Yasmeen Shorish, "Ethics in Research Use of Library Patron Data: Glossary and Explainer," Digital Library Federation. Dec 13 2018, doi:10.17605/OSF.IO/XFKZ6.